\documentclass[12pt]{article}
\newwrite\ffile\global\newcount\figno \global\figno=1

\def\writedef#1{}

\input epsf
\def\figin{\epsfcheck\figin}\def\figins{\epsfcheck\figins}
\def\epsfcheck{\ifx\epsfbox\UnDeFiNeD
\message{(NO epsf.tex, FIGURES WILL BE IGNORED)}
\gdef\figin##1{\vskip2in}\gdef\figins##1{\hskip.5in}% blank space instead
\else\message{(FIGURES WILL BE INCLUDED)}%
\gdef\figin##1{##1}\gdef\figins##1{##1}\fi}

\def\figinsert{}
\def\ifig#1#2#3{\xdef#1{fig.~\the\figno}
\writedef{#1\leftbracket fig.\noexpand~\the\figno}%
\figinsert\figin{\centerline{#3}}\medskip\centerline{\vbox{\baselineskip12pt
\advance\hsize by -1truein\center\footnotesize{ Fig.~\the\figno.} #2}}
\bigskip\endinsert\global\advance\figno by1}
\def\endinsert{}

\begin{document}
\baselineskip 18pt
\newcommand{\Tr}{\mbox{Tr\,}}
\newcommand{\beq}{\begin{equation}}
\newcommand{\eeq}{\end{equation}}
\newcommand{\bea}{\begin{eqnarray}}
\newcommand{\eea}{\end{eqnarray}}
\renewcommand{\Re}{\mbox{Re}\,}
\renewcommand{\Im}{\mbox{Im}\,}
\begin{titlepage}

\begin{picture}(0,44)(0,0)
\put(350,54){SHEP 0321}
\put(350,37){CPHT-RR 043.0703}
\put(350,20){Bicocca-FT-03-22}
\put(350,3){IFUP-TH/2003/24}
\end{picture}
%\hfill SHEP \par
%\hfill CPHT-RR 043.0703 \par
%\hfill Bicocca-FT-03- \par
%\hfill IFUP-TH/2003/
%\par\hfill
\\
\vskip .1in \hfill hep-th/0308006
\hfill
\begin{center}
\hfill
\vskip .35in
{\large\bf
%
%\title
Confinement, Glueballs and Strings from Deformed AdS}
%\maketitle
\end{center}
\vskip .4in
\begin{center}
{\large Riccardo Apreda${}^{a,b,c}$, David E. Crooks${}^{c}$, Nick Evans${}^{c}$ and Michela Petrini${}^{d}$}
%\renewcommand{\footnoterule}{\vspace{-2cm}}
%\footnotetext
%\footnotetext{
%e-mail: \ \ apreda@df.unipi.it, \ \ dc@hep.phys.soton.ac.uk, \ \ \\
%}}
%\hspace*{19.6mm}
%\put(160,-285){\makebox(0,0)[b]{
%evans@phys.soton.ac.uk, \ \
%Michela.Petrini@cpht.polytechnique.fr
%}
\vskip .1in
{\em a Dip. di Fisica, Universit{\`a} di Milano-Bicocca, P. della Scienza, 3; I-20126 Milano, Italy.}\\
{\em b Dipartimento di Fisica, Universit{\`a} di Pisa, via Buonarroti, 2; I-56127 Pisa, Italy.}\\
{\em c Department of Physics, Southampton University, Southampton,
S017 1BJ, UK}\\
{\em d Centre de Physique Th{\'e}orique, {\`E}cole Polytechnique;
F-91128 Palaiseau Cedex, France.}\\

\end{center}
\vskip .33in
\begin{center} {\bf ABSTRACT} \end{center}
\begin{quotation}
\begin{picture}(0,0)(0,0)
\put(-47,-253){\line(1,0){195}}
\put(96.3,-266){\makebox(0,0)[b]{
%\footnotetext{
{\footnotesize
e-mail: \ \ apreda@df.unipi.it, \ \ dc@hep.phys.soton.ac.uk, \ \
}}}
%\hspace*{14mm}
\put(153,-278){\makebox(0,0)[b]{
\footnotesize{evans@phys.soton.ac.uk, \ \
Michela.Petrini@cpht.polytechnique.fr}
}
}
%\vfill
\end{picture}
\noindent We study aspects of confinement in two deformed versions
of the AdS/CFT correspondence - the GPPZ dual of ${\cal N}=1^*$
Yang Mills, and the Yang Mills$^*$ ${\cal N}=0$ dual. Both
geometries describe discrete glueball spectra which we calculate
numerically. The results agree at the 10$\%$ level  with previous
AdS/CFT computations in the Klebanov Strassler background and AdS
Schwarzchild respectively. We also calculate the spectra of bound
states of the massive fermions in these geometries and show that
they are light, so not decoupled from the dynamics. We then study
the behaviour of Wilson loops in the 10d lifts of these
geometries. We find a transition from AdS-like strings in the UV
to strings that interact with the unknown physics of the central
singularity of the space in the IR.
\end{quotation}
\vfill
\end{titlepage}
\eject
\noindent

\section{Introduction}

The AdS/CFT correspondence \cite{mald,kleb,ed1}
has led to the development of dual string theory (supergravity) descriptions
of a variety of large $N$ gauge theories.
As originally stated the duality applied to gauge theories with maximal
supersymmetry including, in the case of four
dimensions, ${\cal N}=4$ Super Yang
Mills. An interesting task is then to
try to move in the direction of gauge theories which are more
phenomenologically applicable. This necessarily
means looking at cases with less than the maximal supersymmetry
of the original duality.
Several examples of
such dual geometries have been constructed including the non-supersymmetric
AdS Schwarzchild geometry \cite{ed2}, and the ${\cal N}=1$ geometries
of Klebanov
Strassler \cite{ks} and Maldacena Nunez \cite{mn}. It is now well understood
how to compute
the glueball bound state predictions \cite{coot} and Wilson loop
behaviour \cite{mloop}
in such
theories and these geometries show confining behaviour.

In this paper we will consider some
of the simplest
interesting deformations of AdS, which will provide results for glueballs and
confining strings that test the systematic errors within such AdS approaches.
The geometries we will study are asymptotically AdS with deformations in the
interior that can be interpreted as including relevant (mass) operators in
the gauge theory. Thus the theories are strongly coupled ${\cal N}=4$ super
Yang Mills in the UV but have less symmetry in the IR. The simplest
method for constructing such deformations \cite{gppz1}
is to use five dimensional
${\cal N}=8$ gauged supergravity where the scalar fields act as source and
vev for operators in the field theory. Lifting the solutions back to 10d
\cite{lift} is somewhat involved but the lifts of the two geometries we
will study are
in the literature \cite{pwn=1,bce2}
(in fact the  ${\cal N}=1^*$ lift is not quite complete).

The first of the geometries we study is the GPPZ ${\cal N}=1^*$ solution
\cite{gppz3}  where one of the scalars of ${\cal N}=8$ supergravity
has a non-trivial profile
corresponding to a mass term for three of the four fermionic fields of
${\cal N}=4$ super Yang Mills. The solution is supersymmetric so the
6 scalars are also massive. The IR of the theory has just a gauge field
and a gaugino. Glueballs in such geometries correspond to excitations of
the dilaton field in the background.  The potential for such fluctuations
%in the relevant Schroedinger equation has already been found in
%\cite{pzconf}. The potential
is a  well \cite{pzconf} and therefore gives rise to discrete
solutions, showing that the theory has a mass gap and is confining
- here we numerically compute the solutions and find the mass
spectrum. The results match to two significant figures the
predictions \cite{kras} from the  Klebanov Strassler \cite{ks}
construction, which should describe  a similar IR gauge theory.
This encouragingly suggests that the UV completions of the
theories, which are different, are unimportant to the glueball
spectrum.
%The  GPPZ solutions also include a set of solutions in which
%a real gaugino condensate is switched on.
In the  GPPZ solutions one can also switch on a second scalar
corresponding to a real gaugino condensate \cite{gppz3}. Previous
analysis has suggested that the solutions  with too large a
condensate are unphysical. From the glueball analysis we confirm
the existence of an  upper limit on the gaugino condensate
parameter  since above it the glueball potential becomes unbounded
from below. Within the physical solutions the value of the gaugino
condensate only changes the glueball masses by of order 10$\%$. It
is also straightforward, following the analysis of \cite{gub}, to
calculate the masses of the bound states of ($\lambda \lambda$)
gauginos. Since there are a set of solutions with different
gaugino condensate the gravity dual describes  a moduli space and
therefore we find a massless bound state plus excitations. The
degeneracy of the vacua is a symptom of large $N$ and is
unfortunately unhelpful for comparison to finite $N$ lattice
computations \cite{mont} where there is no massless state. One
might hope to compare our glueball results to future lattice
results though. Finally we test the decoupling of the UV degrees
of freedom by calculating the masses of the bound states of heavy
fermions; we find they are of comparable size to the glueball
masses so the decoupling is far from complete. This makes the
insensitivity of the glueball masses to the UV physics all the
more remarkable. In principle one should take into account mixing
between these states and those we consider. Such an analysis can
be found for the ${\cal N}=1^*$ theory in \cite{hb4}. Comparison
to those results shows that the mixing again only effects the
glueball masse at the 10$\%$ level.

Our second geometry is the Yang Mills$^*$ dual \cite{bce2}
which in five dimensions
has a single scalar switched on, corresponding to an equal mass for all four
fermion fields of ${\cal N}=4$ super Yang Mills. Brane probing the 10d lift
\cite{bce2}
shows that the six scalar fields also have a mass and therefore the IR of the
theory is just a non-supersymmetric gauge field. We find the potential
relevant for the glueball spectrum of the theory, show that it is a
bounded well and compute the discrete glueball spectrum. The spectrum
 matches at the
10$\%$ level with previous estimates of the spectrum from AdS Schwarzchild
\cite{coot}.
This is again a surprising level of insensitivity to the UV completion since
the AdS Schwarzchild geometry describes a 5d gauge theory in the UV whilst
YM$^*$ returns to 4d ${\cal N}=4$ super Yang Mills. We also calculate
the fermionic bound state masses and find they lie close to the glueball
spectrum showing the limitations of the decoupling.

Although these deformed geometries are easy to construct and work
with we must be careful because they are singular in the deep
interior. Naively one would conclude that there is something badly
amiss! However, in a similar construction of the  ${\cal N}=2^*$
theory \cite{pw,bs2,ep,bpp,ejp} the interior is also divergent.
There brane probing has revealed that the singularity matches to
the pole of the one--loop coupling of the gauge theory (the
enhan\c con mechanism \cite{enhan}). This is a possible
interpretation of the singularities here too. It is also possible
that the configurations are analogous to the Polchinski Strassler
${\cal N}=1^*$ construction \cite{ps} in which the core of the
geometry contains some complicated fuzzy expansion of the D3
branes although identifying such an object from the geometry
appears hard. Further one might expect that the strong coupling
might trigger additional operators to switch on and these might
act to smooth the singularity. This is the mechanism that leads to
smooth geometries in the Klebanov Strassler and Maldacena Nunez
geometries. Thus our geometries may well be incomplete.
Nevertheless they do have the correct UV behaviour and conformal
symmetry breaking. We might hope that typical strong coupling
phenomena will be triggered by a large but finite gauge coupling
and that the geometry very close to the singularity will therefore
not play a role in the geometry's predictions. The glueball
computations support this conclusion since the bounded well that
controls their mass forms before the singularity is reached.

To investigate this further we also study Wilson loops in these
geometries. The Yang Mills$^*$ 10d lift is the most complete so we
concentrate on that case. A probe fundamental string in the
geometry is associated with the interaction between a test quark
anti-quark pair in the background gauge configuration
\cite{mloop}. In pure AdS the wider the quarks are separated the
deeper the string penetrates into AdS. The action of the string
looks like a constant (corresponding to the quark mass) plus a
term determined by conformal symmetry that goes as $- 1/L$ where
$L$ is the quark separation. In the Yang Mills$^*$ geometry the
solutions of the equations of motion of the string behave very
differently. Asymptotically the geometry is AdS so for quarks that
are close the behaviour is the same. However, as the string
samples further into the deformed space the Euler Lagrange
equation solutions describe a second solution in which the quarks
are closer together again. We show that this solution has higher
action than the AdS-like solution. Naively there appears to be a
maximum length string (we suggested this interpretation in the
first version of this manuscript). To understand the solutions
better we study the action of a set of curves linking two quarks
at each fixed separation. In fact a simple sine wave configuration
linking the quarks is sufficient to reveal the role of the
different solutions, even though this set of curves only contains
an approximation to the true solution of the Euler Lagrange
equations. We find that the second solution is in fact a global
maximum of the action. Strings that penetrate deeper into the core
than that configuration have falling action and the string in fact
wants to fall into the singularity. We find a  phase transition as
the quarks are separated where AdS-like strings are replaced by
strings entering the core. This is unfortunate since it means the
physics described by the geometry can not be completely described
by just the supergravity. A stringy resolution of the singularity
is needed. One can imagine that all might yet be well since if
there is some fuzzy brane construction then the set up might be
consistent with confinement by mechanisms similar to those
suggested in the Polchinski Strassler ${\cal N} =1^*$ set up
\cite{ps}. Our knowledge is clearly incomplete though and this
should be taken into account when considering the glueball
spectrum results.

We begin in the next section
with a review of the formalism for constructing deformed AdS theories
and looking at fluctuations about such geometries.
In section 3 we will study the glueballs of ${\cal N}=1^*$, and in
section 4 the glueballs of Yang Mills$^*$. In section 5 we turn to
the analysis of Wilson loops in these geometries.

\section{ Deformations and Fluctuations }

We briefly review the formalism for constructing deformed AdS geometries in
5d supergravity and for studying the fluctuations about such backgrounds.

\subsection{Deforming AdS}

The AdS/CFT correspondence \cite{mald,ed1} maps supergravity
fields to operators in the field theory. The theories we consider
here correspond to deformations of ${\cal N}=4$ Super Yang Mills
by mass terms for the four fermions of the theory. These operators
correspond to scalar fields of the 5d supergravity theory. The
dependence of the supergravity field on the  radial AdS coordinate
encodes information about the RG flow of the field theory
operator.

Thus we look for solutions involving only scalar fields and
allow the scalars to vary only in the radial direction of AdS$_5$ \cite{gppz1}.
The other non trivial field is the metric
\beq
ds^2 = e^{2 A(r)}dx^\mu dx_\mu + dr^2,
\label{metric4}
\eeq
where $\mu=0,\ldots,3$, $r$ is the radial direction in AdS$_5$, and
$A(r)\rightarrow r$ as $r \rightarrow \infty$ so that the space
becomes AdS$_5$ at infinity. With an appropriate parametrization the scalar
lagrangian can be written as ${\cal L} = {1 \over 2} (\partial \lambda)^2
- V(\lambda)$ and the
%Thus we study scalar
%fields with lagrangian ${\cal L} = {1 \over 2} (\partial \lambda)^2
%- V(\lambda)$ and allow the scalar
%fields to vary only in the radial direction of AdS$_5$ \cite{gppz1}.
%We parametrize the metric as
%\beq
%ds^2 = e^{2 A(r)}dx^\mu dx_\mu + dr^2
%\eeq
%where $\mu=0..3$, $r$ is the radial direction in AdS$_5$, and
%$A(r)\rightarrow r$ as $r \rightarrow \infty$ so that the space
%becomes AdS$_5$
%at infinity.
equations of motion for $\lambda$ and $A$ are \cite{gppz1}
\beq \label{e1}
\lambda^{''} + 4 A^{'} \lambda^{'} = {\partial V \over \partial \lambda},
\hspace{1cm}
6 A^{'2} = \lambda^{'2} - 2 V.
\eeq

For large $r$, where the solution will return to AdS$_5$ at first
order and $\lambda \rightarrow 0$ and $V \rightarrow {m^2 \over 2} \lambda^2$,
only the first equation survives with solution
\beq
\lambda = ae^{-(4 - \Delta) r} + be^{-\Delta r}
\label{asb}
\eeq
$a$ and $b$ are constants, while the conformal dimension $\Delta$ of the field
theory operator is related to the mass of the supergravity scalar by
\beq
M^2 = \Delta(\Delta-4).
\label{mass}
\eeq
$a$ is interpreted as the source for a field theory operator and $b$
as the vev of that operator, since $e^r$ has conformal dimension one.

If the solution retains some supersymmetry then the potential can
be written in terms of a superpotential \cite{freed1}
\beq
V = { 1 \over 8} \left| { \partial W \over \partial \lambda} \right|^2 -
{1 \over 3} |W|^2
\eeq
and the second order equations reduce to first order
\beq \label{susyeom}
\lambda^{'} = { 1 \over 2} {\partial W \over \partial \lambda}, \hspace{1cm}
A^{'} = - {1 \over 3} W.
\eeq

\subsection{Linearized Fluctuations}

The supergravity scalars characterised by a non zero $b$ in
(\ref{asb}) map to field theory operator expectation
values.  We can therefore look for linearized fluctuations of a
given  scalar in a background, corresponding to bound states in the
field theory  associated with that operator.
We follow the analysis in \cite{ed2,coot,gub}. We look for normalisable
solutions of the linearized wave equation
\beq
\partial_\mu( \sqrt{-g} g^{\mu \nu} \partial_\nu) \delta \Phi
= {\partial ^2 V
\over \partial \Phi \partial \Phi} \delta \Phi,
\label{fluct}
\eeq
that behave  as plane waves at the boundary
\beq
\delta \Phi = \psi(r) e^{-i k x}.
\eeq
To determine the spectrum of the fluctuations, a standard procedure in
AdS is to reduce the above equation to a solution of a Schroedinger
problem \cite{ed2,coot,gub}.
Making the change of coordinates ($r \rightarrow z$)
\beq
{d z \over dr} = e^{-A}
\eeq
and rescaling
\beq
\psi \rightarrow e^{-3 A/2} \psi,
\eeq
equation (\ref{fluct}) takes the Schroedinger form
\beq \label{se}
( - \partial_z^2 + U(z)) \psi(z) = M^2 \psi(z),
\eeq
where
\beq \label{u}
U= {3 \over 2} A^{''} + {9 \over 4} (A^{'})^2 +e^{2 A} {\partial ^2 V
\over \partial \Phi \partial \Phi}.
\eeq

Primes now denote differentiation with respect to $z$. If this potential
takes the form of a bounded well then there are discrete solutions
(with $\ k^2 = -M^2$)
for the linearized fluctuation indicating confinement of the fields
into discrete bound states.

Since we will work throughout in the $z$ coordinates, we re-express the
results of the previous subsection in these new  coordinates.
The second order equations of motion for the background
deformation become
\beq
\lambda^{''} + 3 \lambda^{'} A^{'} = e^{2 A} {\partial V
\over \partial \lambda} , \hspace{1cm}
6 A^{'2} = \lambda^{'2} - 2 e^{2 A} V.
 \eeq
In the UV ($z \rightarrow 0$) the behaviour is
\beq \label{lamuv}
z=-e^{-r}, \hspace{1cm} e^{2A} = {1 \over z^2}, \hspace{1cm}
\lambda = a (-z)^{4- \Delta}+ b (-z)^\Delta,
\eeq
where, again, $a$ is interpreted as a source for an operator and $b$
as the vev of that operator since $z$ has conformal dimension -1.
Finally the first
order equations (\ref{susyeom}) now read
\beq \label{susyeomz}
\lambda^{'} = { 1 \over 2} e^{A} {\partial W \over \partial \lambda},
\hspace{1cm}
A^{'} = - e^{A}{1 \over 3} W.
\eeq
\newpage

\section{The ${\cal N}=1^*$ Geometry}

The ${\cal N}=1^*$ theory is ${\cal N}=4$ super Yang Mills with equal mass terms for
the three adjoint chiral superfields leaving just the vector multiplet
massless. The UV of the theory is conformal whilst the IR behaves like
${\cal N}=1$ Yang Mills generating a gaugino condensate dynamically
in the weak coupling limit. At large $N$,
where we will be working,  the UV is also strongly coupled so the
chiral multiplets can not be considered fully decoupled. The
 theory at large $N$ has been studied in \cite{dor} and
it has been shown to have a set
of discrete vacua differentiated by the magnitude of the gaugino condensate
which is real.

The gravity dual of this theory was found by GPPZ \cite{gppz3}.
%In the 5d supergravity theory we identify
%the scalars
%$m$ and $\sigma$ from the 10 of the SO(6) gauge symmetry of AdS$_5$ (the
%SU(4)$_R$ symmetry of the gauge theory) dual to the dimension 3
%operators
%\beq \label{ms}
%{\cal O}_{m} = \sum_{i=2}^4 \psi_i \psi_i, \hspace{1cm} {\cal O}_{\sigma}
%= \psi_1 \psi_1.
%\eeq
In the 5d supergravity theory
one must identify the scalars dual to the
the dimension 3 operators
\beq \label{ms}
{\cal O}_{m} = \sum_{i=2}^4 \psi_i \psi_i, \hspace{1cm} {\cal O}_{\sigma}
= \psi_1 \psi_1.
\eeq
These corresponds to two scalars $m$ and $\sigma$ in the 10 of the SO(6)
gauge symmetry of the supergravity theory
(the global SU(4)$_R$ symmetry of the gauge theory).

The GPPZ flow tuned by these two scalars preserves an $SO(3)$ flavour symmetry.
%The full field content of the supergravity model is given by
The supergravity fields that are singlet under this residual $SO(3)$ can be
organized in one gravity multiplet and two other hypermultiplets \cite{pwn=1}.

The potential for $\sigma$ and $m$ is \cite{gppz3}
\beq \label{n1v}
V=-\frac{3}{8} [\cosh^2(\frac{2m}{\sqrt{3}})+4 \cosh(\frac{2m}{\sqrt{3}})
\cosh(2\sigma)-\cosh^2(2\sigma)+4],
\eeq
and can be obtained from a superpotential
\beq
W=-\frac{3}{4}\left[\cosh(\frac{2m}{\sqrt{3}})+\cosh(2\sigma)\right],
\eeq
as expected since the solutions should maintain ${\cal N}=1$ supersymmetry.
We can then work with the first order equations
%Working with the first order equations  ensures that N=1 supersymmetry is
%preserved and that the flow equations reduce to
\bea \label{n1*}
\frac{\partial \sigma}{\partial z}&=&-\frac{3}{2}e^{A} \sinh (2 \sigma), \\
\frac{\partial m}{\partial z}&=&-\frac{\sqrt{3}}{2}e^{A} \sinh (\frac{2m}{\sqrt{3}}), \\
\frac{\partial A}{\partial z}&=&\frac{1}{2}e^{A} [\cosh(\frac{2m}{\sqrt{3}})+\cosh (2 \sigma) ].
\label{n12*}
\eea
Equations (\ref{n1*}-\ref{n12*}) can be solved analytically \cite{gppz3}, and the result
is a family of solutions parametrized by the boundary
values of the fields as $z\rightarrow 0$ (UV)
\beq
m \rightarrow -a z , \hspace{1cm}
\sigma \rightarrow -b z^3  , \hspace{1cm}
A \rightarrow - \log |z|.
\label{bc}
\eeq
From (\ref{lamuv}), $m$ corresponds to a mass for three of the adjoint fermions
and $\sigma$ is the  gaugino condensate.
In this paper we do not need the analytic expression for the flows, so
we will solve the first order equations numerically subject
to the boundary conditions (\ref{bc}).

Solutions exist for all initial choices of $a$ and $b$ but the
resulting geometries are singular. In \cite{gub}
it was argued that only a
subset of these flows should be considered physical. There it was required that
the supergravity potential evaluated on the solution should be bounded
from above. This condition restricts us to flows with
initial conditions
\bea \label{bound}
b &\leq& 3^{-3/2} a \\
&\approx& 0.19 a \nonumber
\eea
The gravity dual therefore predicts a moduli space of vacua with
varying real gaugino condensate up to some maximum value ($0.19m$)
which is plausibly the $N \rightarrow \infty$
limit of the field theory result discussed above.

We will see further evidence for this bound on $a$ shortly.
For convenience in what follows we will set $a=1$ - that is we set the scale of
all mass operators in terms of the chiral multiplets' mass.

\subsection{Glueballs}

The mass spectrum of bound states can be obtained from fluctuations of
the dual scalar fields by finding the eigenvalues
of (\ref{se}). We first calculate the mass spectrum of glueballs with
quantum numbers $J^{PC}=O^{++}$ associated with the
correlators $\langle \Tr F^2(x) \Tr F^2(y) \rangle$.
The operator ${\cal O} = \Tr F^2$ is dual to the dilaton, so we need the
Schroedinger potential (\ref{u}) in this case. The dilaton
does not contribute to the supergravity scalar potential, so that the
Schroedinger potential for the glueballs is just
\beq \label{gpot}
U= {3 \over 2} A^{''} + {9 \over 4} (A^{'})^2.
\eeq

Actually the dynamics of the eight scalars of the full
supergravity background is quite complicate and the dilaton,
though being inert (that is having no radial dependence in the
background solution) couples to the other scalars. To find the
complete Schroedinger-like equation for the dilaton one has first
to diagonalize the equation of motions for the system of scalars,
as in \cite{anatomy}, where an analytic solution has been found
for all the scalars. Bound state spectra for some scalars had been
previously calculated also in
\cite{boupre,boupre1,boupre2,hb1,hb4}.

In order to obtain a discrete spectrum with a mass gap $U$ must be bounded
below. We plot the potential for varying $b$ in Figure 1.
The potential  is indeed bounded for all
physical flows satisfying the bound (\ref{bound}).
The unphysical flows with $b > 1.9 a$, which violate (\ref{bound}),
have an unbounded glueball potential. Thus we have further evidence for
the bound.

The eigenvalues of (\ref{se}), and hence the glueball mass spectrum, can
be easily obtained using the numerical shooting
method. Since we want the solution to match to the operator $\Tr F^2$ we
set the UV ($z\rightarrow 0$) boundary conditions on $\psi$ to be
\beq
\psi(z) = {1 \over z^4},
\eeq
and then numerically solve for various $M$ seeking regular solutions
in the IR.

The results for $b=0$ and $b=0.19$ are shown in Table 1.
The $0^{++}$ mass is not a prediction, but just sets the strong coupling
scale, so we normalize the lightest glueball state to 1. The
gaugino condensate's value has little influence on the glueball masses
(at most of order 10$\%$).

We also quote the glueball mass spectrum obtained from the ${\cal
N}=1$ Klebanov Strassler model \cite{ks,kras}. The Klebanov
Strassler dual also describes ${\cal N}=1$ Yang Mills in the IR
but in the UV there are extra flavours that participate in a
cascade of Seiberg dualities. The glueball predictions agree very
well with the ${\cal N}=1^*$ results, suggesting the UV completion
of the theory is relatively unimportant to the glueball spectrum.
The Maldacena Nunez \cite{mn} dual also describes ${\cal N}=1$
Yang Mills in the IR and the glueball spectrum has been recently
analyzed in \cite{apt}. That dual though becomes strongly coupled
in the UV so the computation is less clear. In  \cite{apt} a UV
cut off is imposed and then the glueball masses calculated - they
are rather dependent on the choice of cut off. As noted by those
authors the agreement with the Klebanov Strassler results is not
so good.

\begin{center}
\hskip-10pt{\lower15pt\hbox{
\epsfysize=2.5 truein \epsfbox{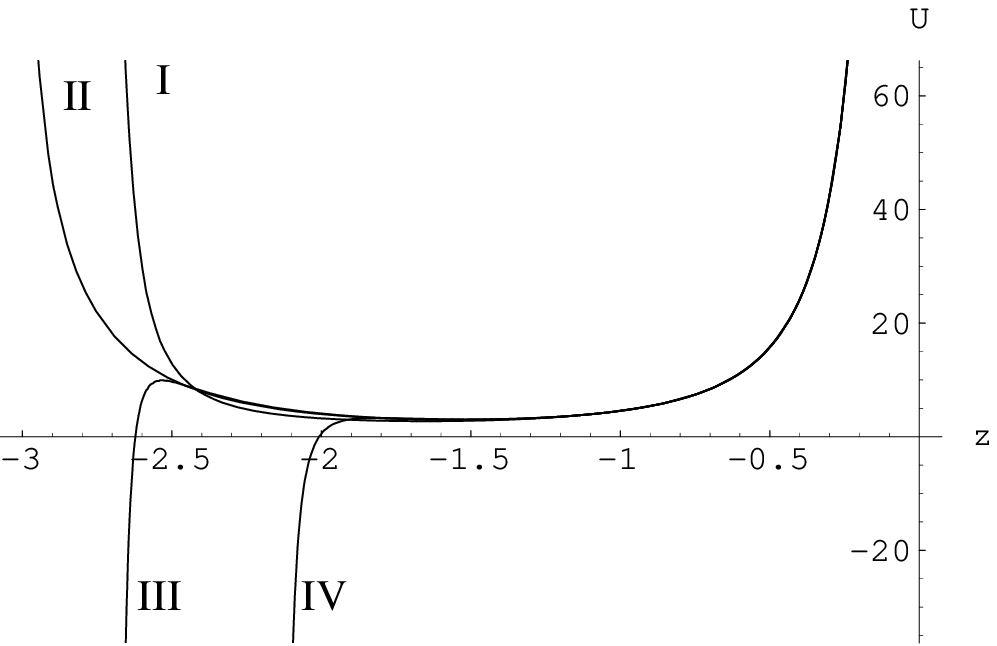}}}\medskip\\[5mm]
Figure 1:The glueball potential in ${\cal N}=1^*$ with $a=1$ and: I) $b=0.1$;
II) $ b=0.19$; III) $b=0.2$; IV) $b=0.25$.
\end{center}
\vspace*{5mm}

\begin{center}
\begin{tabular}{|c||c|c|c|c|} \hline
State & ${\cal N}=1^*$ ($b=0$) & ${\cal N}=1^*$ ($b=0.19$)
& ${\cal N}=1$ KS & $(pL)^2 4 (n+1) (n+4)$
\\
\hline \hline
$0^{++}$ & 1.0 (input)& 1.0 (input) & 1.0 (input) &  1.0 (input) \\
\hline
$0^{++*}$ & 1.5 & 1.5 & 1.5 & 1.6 \\
\hline
$0^{++**}$ & 2.0 & 1.9 & 2.0 & 2.1\\
\hline
$0^{++***}$ & 2.5 & 2.3 & 2.5 & 2.6\\
\hline
$0^{++****}$ & 3.0 & 2.7 & 3.0 & 3.2\\
\hline
\end{tabular}\medskip
\\[2mm]
Table 1: Spectrum of ${\cal N}=1^*$ glueball masses from
supergravity with $b=0$ and $b=0.19$, not taking into account the
mixing. In the next column the ${\cal N}=1$ glueball spectrum
obtained from Klebanov Strassler background is shown for
comparison. The last column  shows the analytic result obtained in
\cite{hb4} taking into account mixing. In all cases the
$0^{++}$ mass has been scaled to 1.
\end{center}

\subsection{Gluino Bound States}

We now move on to investigate the masses of bound states of the
fermions of the ${\cal N}=1^*$ theory. For simplicity we will set
$\sigma = 0$ ($b=0$) in what follows. To study bound states of the
massless gluino we look at fluctuations of the scalar $\sigma$.
The corresponding potential (\ref{u}) is \beq U_{\sigma}= {3 \over
2} A^{''} + {9 \over 4} (A^{'})^2 +3 e^{2 A} \left[1-2
\cosh(\frac{2 m}{\sqrt{3}})\right]. \eeq

Numerically plotting this potential reveals it to be bounded giving a
discrete spectrum with a mass gap. As in the case of the glueballs, we
can find the
eigenvalues of (\ref{se}) with the shooting method. The results are
shown in Table 2 below, normalized to the lightest glueball mass above.
The gravity dual describes a moduli
space of vacua corresponding to different background values of $\sigma$
so we would expect to find a massless boson. In fact we do and the smallest
eigenvalue is zero.

Considerable work has been done by the DESY collaboration \cite{mont} studying
pure ${\cal N}= 1$ Yang Mills on the lattice. They reach the supersymmetric
theory by tuning the gaugino mass to zero and to date their results
for bound state masses are still somewhat away from the supersymmetric point.
In the future though it should be possible to extract the supersymmetric
values and it would be nice to have AdS predictions available for comparison.
The lattice theory at finite $N$ has an anomalous U(1)$_R$ symmetry and
a complete mass gap. On the other hand
in the ${\cal N}= 1^*$ theory that U(1)$_R$ symmetry is only an accidental
symmetry in the IR in the weak coupling limit and is not present at all in the
large $N$ theory \cite{dor} plus the gravity dual describes a
massless bound state.
The fermionic vacuum structure is therefore
quite different and it is hard to compare the results for
fermionic bound states. The best hope would be to compare
the 0$^{++}$ glueball spectrum above.

Finally it is also interesting to look at bound states of the massive
fermions to see how decoupled these fields are from the dynamics.
There is an SO(3) symmetry acting on the massive fermions so it is sufficient
to analyse bound states of a single massive flavour. We therefore subdivide
the operator ${\cal O}_{m}$ further introducing the
scalars $\mu$ and $\nu$ corresponding to
\beq \label{opm}
{\cal O}_{\mu} = \psi_2 \psi_2, \hspace{1cm} {\cal O}_{\nu}
= \sum_{i=3}^4 \psi_i \psi_i.
\eeq

This enables us to study bound states of the massive fermion
$\psi_{2}$ by looking at fluctuations of $\mu$. In terms of
these scalars the potential is \cite{ep}
\beq \begin{array}{ccl}
V& = & \frac{1}{8} \left[ -5+\cosh(4\mu)- 4\cosh(2\mu)\right]
-\cosh(\sqrt{2}\nu) [\cosh(2 \mu)+1] \\
&&\\
&&+\frac{1}{16}\left[-3 + 2 \cosh (2
\sqrt{2} \nu) + \cosh (4 \mu)\right] \end{array}.
\eeq

Of course, we are still interested in preserving ${\cal N}=1$
supersymmetry in the background. If we set \beq \nu =
\sqrt{\frac{2}{3}} m , \hspace{1cm} \mu= \frac{m}{\sqrt{3}} \eeq
then we recover the ${\cal N}=1^*$ potential with $\sigma=0$. The
Schroedinger potential we are interested in for fluctuations in
$\mu$ about this background is
\beq U_{\mu}= {3 \over 2} A^{''} +
{9 \over 4} (A^{'})^2 + e^{2 A} \left[-2-2 \cosh(\frac{2
m}{\sqrt{3}})+\cosh(\frac{4m}{\sqrt{3}})\right]. \eeq This is
indeed a bounded potential. The mass spectrum is found by shooting
and the results are displayed in Table 2. We find that the fermion
bound states are a little heavier than the lightest glueball but
far from completely decoupled.

Since these states are not decoupled we must worry about mixing
between them and the states whose masses we have calculated above.
In \cite{hb1,hb4, anatomy} the authors make progress in this
direction by calculating the spectra through the analysis of the
two point correlation function for the associated operator, using
the holographic renormalization method \cite{hb2,hb3,hb4}. The
glueball spectrum associated to the dilaton in the ${\cal N}=1^*$
background was found analytically in \cite{anatomy} solving the
full equations of motion, that is taking into account the mixing
of the dilaton with other scalars in the supergravity theory. The
authors found a spectrum given by $m^2= (pL)^2 4 (n+1) (n+4),$
with $n=0,1,2, \ldots$. This spectrum, indicated in Table 1 above,
is consistent at the $5$-$10\%$ level accuracy with our numerical
result. Our results for the gaugino bound state agree perfectly
with their analytic expression for the spectrum $m^2= (pL)^2 4
(n+1) (n+2)$ with $n=0,1,2,...$. It is remarkable that the physics
of the model is not dramatically affected by the mixing, and
supports a more naive analysis where these extended super-partner
states are neglected.

It is interesting to note that if we study
fluctuations of the supergravity scalar $m$ which corresponds to a composite
operator of all three fermions (\ref{ms}) we obtain the potential
\beq
U_{m}= {3 \over 2} A^{''} + {9 \over 4} (A^{'})^2 +e^{2 A} \left[
-2 \cosh(\frac{2 m}{\sqrt{3}})- \cosh(\frac{4 m}{\sqrt{3}})\right].
\eeq
This potential is unbounded. This presumably corresponds to an instability
in the field theory for such a bound state to decay to the states we have seen
above.\\[2mm]

\begin{center}

\begin{tabular}{|c||c|c|} \hline
State & $\psi_1$ & $\psi_2$ \\
\hline \hline
1 & 0.0& 1.2 \\
\hline
2 & 1.0 & 1.7 \\
\hline
3 & 1.6 & 2.2 \\
\hline
4 & 2.1 & 2.8 \\
\hline
5& 2.7 & 3.3
\\  \hline
\end{tabular}
\\[5mm]
Table 2: The first five bound states of the massless gluino $\psi_{1}$
and the massive fermion $\psi_{2}$ in the ${\cal
N}=1^*$ theory.
\end{center}

\section{The Yang Mills$^*$ Geometry}

The Yang Mills$^*$ theory \cite{bce2} is
${\cal N}=4$ super Yang Mills with mass terms for
all the adjoint matter fields leaving just non-supersymmetric gauge fields
in the IR. At large $N$ the massive fields can again not be considered totally
decoupled.

The Yang Mills$^*$ geometry is obtained by turning on  a supergravity scalar
$\lambda$ from the 10 of SO(6) dual to the operator
\beq \label{opym}
{\cal O} = \sum_{i=1}^4 \psi_i \psi_i.
\eeq
The potential for the scalar can be obtained from the ${\cal N}=1^*$ solution
\cite{gppz3}  by setting the two scalars equal\footnote{To be precise
one must set $m = \sqrt{3/ 4} \lambda$ and $\sigma = \sqrt{1 / 4}
\lambda$ to maintain a canonically normalized kinetic term}:
\beq \label{pot}
V = - {3 \over 2} \left( 1 + \cosh^2 \lambda \right).
\eeq
In this case $M^2=-3$ in (\ref{mass}) and the ultra-violet solutions are
\beq
\lambda = {\cal M} z + {\cal K} z^3.
\eeq

%The field theory
%operators have dimension 1 and 3.
The field theory
mass term  has dimension 3.
Thus in what
follows ${\cal M}=0$ corresponds to a solution with just bi-fermion
vevs while ${\cal K}=0$ corresponds to the purely massive case.
Giving a mass to all four fermions breaks supersymmetry
completely so we would expect the scalars to radiatively acquire masses
so that the deep infra-red should be pure Yang Mills. The analysis of
the 10d lift in \cite{bce2} supports this hypothesis.

\subsection{Numerical Solutions}
We are not able to write down first order equations for the deformations
as there is no supersymmetry and no
superpotential.  Thus we are forced to solve the second
order equations numerically. Numerical solutions of the flow equations
for the scalars $\lambda(z), A(z)$ are displayed in
Figure 2 for different asymptotic boundary conditions. The mass only flow
is unique, lying on the border between two separate behaviour flows.
If there is even a small condensate
$\lambda(z)$ diverges very rapidly. It is necessary to fine tune the
initial conditions to high precision in
order to isolate the mass only flow. All these flows are singular including the
mass only solution.

We need a criteria for deciding which of these flows is the physical
flow for the YM$^*$ theory. The simplest criteria we have found is the
boundedness of the $0^{++}$ glueball Schroedinger potential. Thus we again
look at linearized dilaton fluctuations and evaluate the Schroedinger potential
in (\ref{gpot}). We plot the potential for a variety of flows with both mass
and condensate present in Figure 3.  The presence
of a condensate leads to an unstable geometry, and an unbounded
glueball potential. For the mass
only solution, however, the glueball potential is a bounded well and
the spectrum is calculable. We therefore conclude that this is the physical
flow.

In \cite{gub} it was proposed that a necessary condition for singular
flows to be physical is that the supergravity
potential, evaluated on the solution of the equations of motion,
should be bounded above. This condition is the origin
of the bound (\ref{bound}). Figure 4 shows the supergravity potential
evaluated for different asymptotic boundary
conditions. All the Yang Mills$^*$ flows satisfy this condition,
including the unphysical flows with a condensate. It
is also a necessary condition for a confining gauge theory
that the glueball potential be a bounded well. Here
this appears to be a stricter condition that successful distinguishes
the physical from the unphysical flows.\\[3mm]
\vfill

\begin{center}
\hskip-10pt{\lower15pt\hbox{ \epsfysize=2. truein
\epsfbox{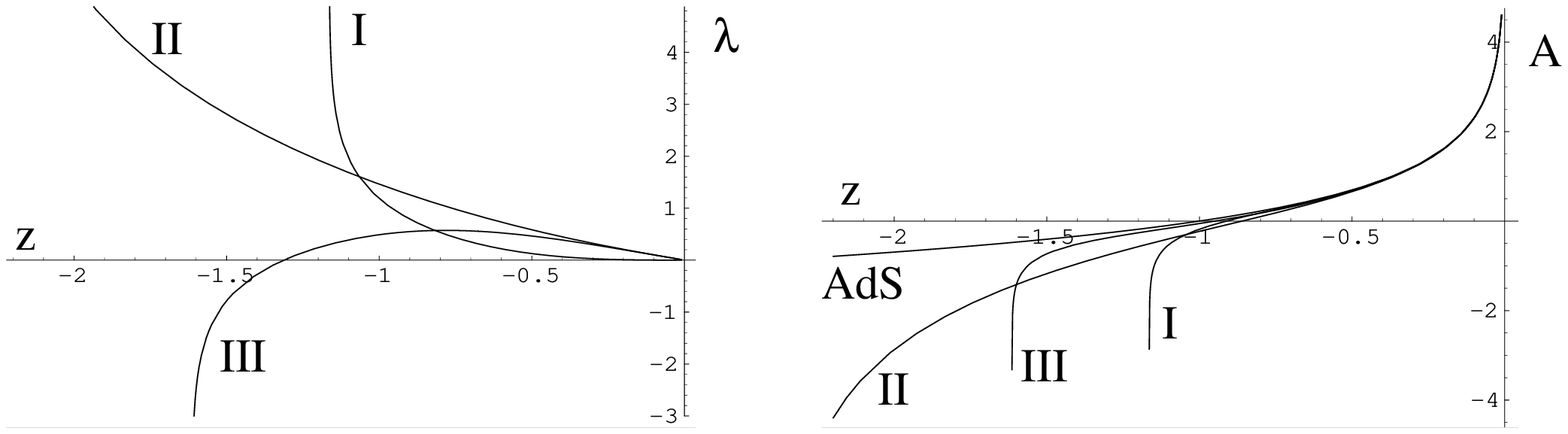}}}\medskip

Figure 2:
Plots of $\lambda$ and $A$ vs $z$ in the YM$^*$ set up
for a variety of UV initial
conditions: \\I) condensate only ($\lambda \simeq -z^3$); II) mass only
($\lambda \simeq -z$); III)
 mass and condensate ($\lambda \simeq -(z + z^3)$).\\ The AdS function $A$ is
also plotted.\\[5mm]
\end{center}

\vfill

\begin{center}
\hskip-10pt{\lower15pt\hbox{ \epsfysize=2. truein
\epsfbox{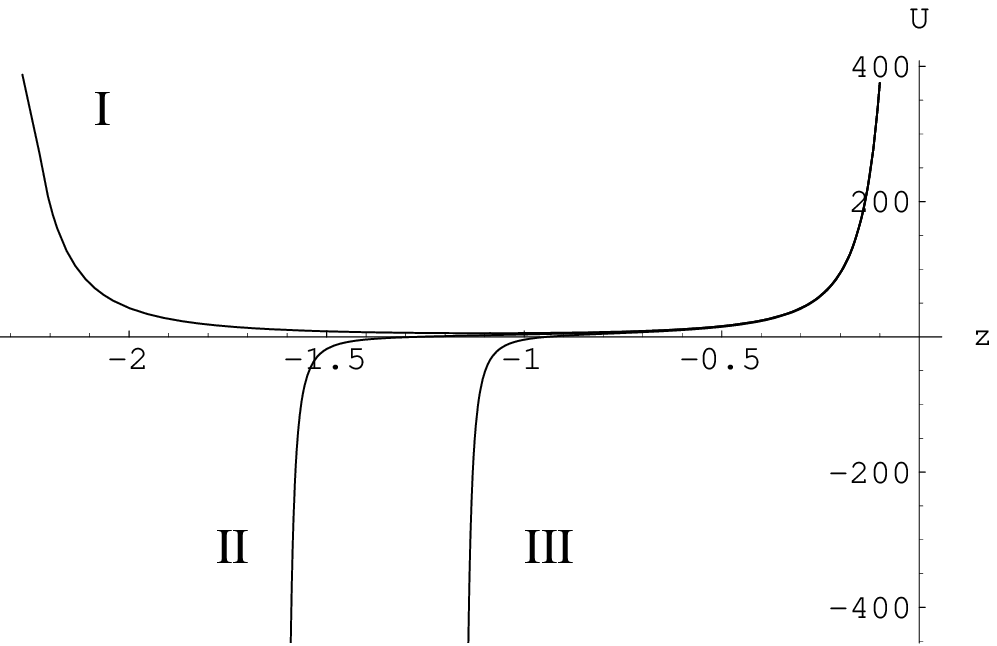}}}\medskip

Figure 3: The glueball potential showing: I) mass only; II) mass and
condensate; III)  condensate only.\\[5mm]
\end{center}
\vfill
\begin{center}
\hskip-10pt{\lower15pt\hbox{ \epsfysize=2. truein
\epsfbox{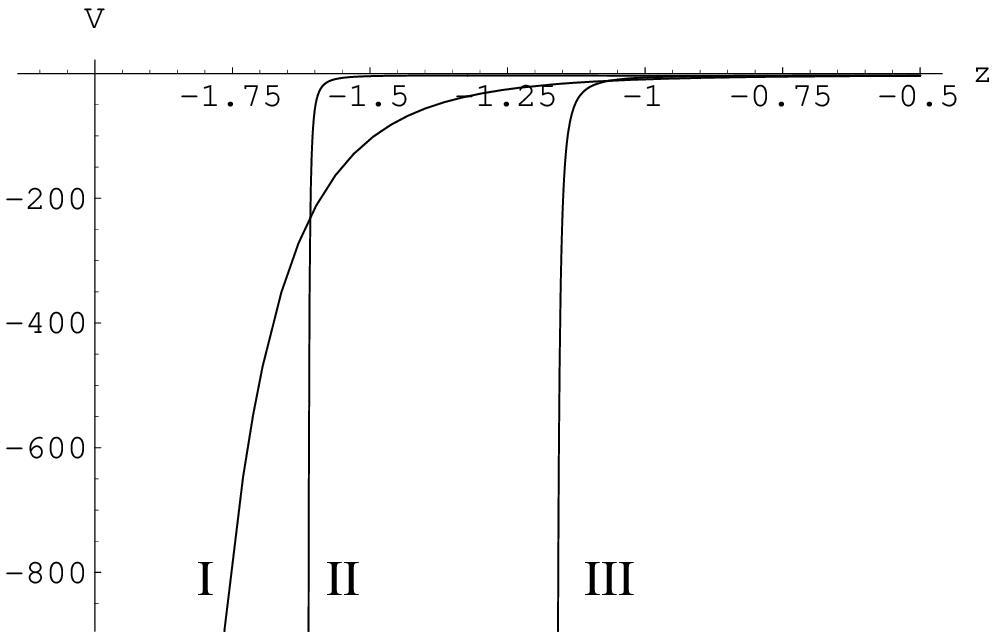}}}\medskip

Figure 4: The supergravity potential for a range of flows: I) mass only;
II) mass and condensate; III)
 condensate only. All cases are bounded above.
\end{center}

\subsection{IR Asymptotic Solutions}

We have also been able to analytically isolate the IR behaviour of the
unique YM$^*$ flow that gives rise to a bounded glueball potential.
In the $r$ coordinates, the infra-red corresponds to
$r \rightarrow r_0$, $\lambda(r)
\rightarrow \infty$ and $A(r) \rightarrow -\infty$. The flow
equations become
\beq
\lambda^{''} + 4 A^{'} \lambda^{'} = -\frac{3}{4} e^{2 \lambda}, \hspace{1cm}
6 A^{'2} = \lambda^{'2} +\frac{3}{4} e^{2 \lambda},
\eeq
and have  solutions
\beq
\lambda =  \log \left(c \over {|r-r_{0}|} \right),
\hspace{1cm} c=\sqrt{\frac{20}{9}},\hspace{1cm} A =\frac{2}{3} \log|r-r_{0}|,
\eeq
where $r_0$ marks the position of the singularity.
This is clearly not a complete set of solutions, so could represent any of the
solutions in Figure 2. To check whether this flow really corresponds to
the solution with just a mass term, we can compute the glueball potential.
This  can easily done by changing variables
$dz=e^{A}dr$
%and computing the glueball potential $U(z)$
\bea
z&=&\int e^{-A} dr \\ \nonumber
 z-z_{0}&=&3(r-r_{0})^{1/3}.
\eea
Hence
\beq
A(z)=2 \log|\frac{z-z_{0}}{3}|,
\eeq
and the glueball potential $U(z)$ in the infra-red is given by
\bea
U(z)&=& {3 \over 2} A^{''} + {9 \over 4} (A^{'})^2 \nonumber \\
&=& \frac{6}{(z-z_{0})^2}.
\eea

We see that $U(z) \rightarrow \infty$ in the infra-red.
Comparison with numerical study of the full second order
equations implies that these solutions are the IR limit of the
mass only Yang Mills$^{*}$ geometry. If there had been a
condensate present, we would have found $U(z) \rightarrow -\infty$.

\subsection{Glueballs}

As discussed above, the $0^{++}$ glueball Schroedinger potential for the
mass only case is a bounded well. We calculate the glueball mass
spectrum by shooting, and the results are shown in Table 3.
As in the previous example, the lightest glueball state is not a
prediction, but can be
used to fix the scale $\Lambda_{QCD}$. As in the ${\cal N}=1^*$
case we normalize the lowest state to 1.

Again a possible source of inaccuracy in the nuerical results can derive from the mixing of the dilaton with other scalars.
Even if there is no analytic proof for such mixing, we expect it to occur by analogy with the supersymmetric case.

An AdS approach to non-supersymmetric Yang Mills glueballs has already
been presented  in the literature.
The AdS Schwarzchild black hole background describing M5 branes
with a compact time
dimension describes the high temperature limit of a strong coupled
5d field theory \cite{ed1,ed2}. The effect of the temperature is to give masses
to all fields except the spatial components of the gauge field. Thus in the
deep IR this theory is a 4d Yang Mills theory. This geometry was used
to compute the glueball spectrum in \cite{coot} and
we display the results in Table 2
for comparison (an alternative calculation of the same spectrum is
given in \cite{newglue}).
The results are surprisingly close, differing only at the
$10\%$ level.
We believe that taking into account the mixing would increase the masses, as happens in the supersymmetric case, making the agreement even more good and more remarkable.
Our theory is 4d ${\cal N}=4$ Yang Mills in
the UV whilst the AdS Schwarzchild dual describes a higher dimension theory
in the UV. Since the UV is not decoupled from the strong interactions
we might have expected at least order one differences. The closeness
of the results hints at the relative stability of glueball masses across
a wide range of theories and strengthens the case for trusting the
AdS method of computation.

It is also interesting to compare the results to lattice computations
in pure Yang Mills.
Data is only available for the lightest two $0^{++}$ states. In Table 3
we show the data for $N=3$ \cite{tmp} and also the $N \rightarrow \infty$
extrapolation of \cite{teper}. The lattice errors are of order 10$\%$.
The AdS results actually match better to the $N=3$ results but still agree
with the $N \rightarrow \infty$ result at the 15$\%$ level. The
data is very limited but again the
agreement is surprising given that the UV of these theories are all so
different.
%- one might have expected order one differences.
\\[3mm]

\begin{center}

\begin{tabular}{|c||c|c|c|c|} \hline
State & Yang Mills* & AdS-Schwarz & Lattice ($N=3$)&Lattice ($N=\infty$)
\\
\hline \hline
$0^{++}$ &  1.0 (input) & 1.0 (input)&  1.0 (input) &  1.0 (input) \\
\hline
$0^{++*}$  & 1.5 & 1.6 & 1.5 &1.9\\
\hline
$0^{++**}$  & 1.9 & 2.1 & -& -\\
\hline
$0^{++***}$  & 2.3 & 2.7 & -& -\\
\hline
$0^{++****}$  & 2.7 & - & -& -\\
\hline
\end{tabular}
\\[5mm]
Table 3: Spectrum of glueball masses from the lattice, from Yang Mills$^*$
and Witten's AdS-Schwarzchild dual. Again, the
lowest glueball mass has been scaled to one in all cases.
\end{center}

\subsection{Fermion Bound States}

To test the degree to which the massive fermions of the YM$^*$ theory have
decoupled we will calculate the mass spectrum of bound states for them.
We use the ${\cal N}=1^*$ potential in ({\ref{n1v}) where the scalar $\sigma$
describes the bi-fermion
operator of a single flavour of adjoint fermions. There is again an SO(4)
symmetry on the massive fermions so it is sufficient to study one flavour.
%with the bi-fermion
%operator of a single flavour of adjoint fermions described by
%the scalar $\sigma$.
Setting the fields $m,\sigma$ to their YM$^*$
background values we then look at fluctuations of $\sigma$.
The appropriate potential (\ref{u}) for this scalar
in the Yang Mills$^*$ background is
\beq
U_{\sigma}= {3 \over 2} A^{''} + {9 \over 4} (A^{'})^2 -3 e^{2 A}.
\eeq

This is again a bounded well potential and the spectrum of bound states
is calculable. The results are shown
in Table 4. The bound states have masses of order the glueball masses
and so can not be considered truly decoupled. Again this suggests that
the success of the AdS computations of glueball masses to match lattice data
above
indicates the stability of the glueball masses across a range of gauge theories
rather than that the decoupling of extra fields is being achieved.

%The potentials (\ref{u}) for $m$ and $\lambda$ are
%
%\beq
%U_{m}= {3 \over 2} A^{''} + {9 \over 4} (A^{'})^2 -4e^{2 A} \cosh^2 \lambda
%\eeq
%
%\beq
%U_{\lambda}= {3 \over 2} A^{''} + {9 \over 4} (A^{'})^2 -e^{2 A}
%\cosh 2 \lambda
%\eeq
%
% These are both unbounded and it is not possible to calculate a mass spectrum
% in these cases.

\begin{center}

\begin{tabular}{|c||c|} \hline
State & mass

\\
\hline \hline
1 & 0.9 \\
\hline
2 & 1.3 \\
\hline
3 & 1.7 \\
\hline
4 & 2.1 \\
\hline
5 & 2.6 \\
\hline
\end{tabular}
\\[3mm]
Table 4: First five bound states of a massive adjoint fermion in
Yang Mills$^*$.
\end{center}

\bigskip

\section{Wilson Loops}

The quark anti-quark interaction potential
may be studied in  AdS duals \cite{mloop}
by introducing a probe D3 brane into the geometry at some radius
$z_{max}$. Fundamental strings between the probe and the central stack of D3
branes would represent W bosons which transform in the ($N,1$) of the
$SU(N)\times U(1)$ gauge group - we may equally think of these states as
quarks since they are in the fundamental representation of $SU(N)$.
Thus a string attached to the probe with well separated ends
play the role of a quark anti-quark pair with mass of order of the energy scale
determined by $z_{max}$. The action of the string corresponds
to the interaction energy between the pair. To study such a configuration
we necessarily require the 10d lifts of our deformed geometries since the
string lives in 10d. The lift of the Yang Mills$^*$ geometry has been found
in \cite{bce2}  while for the ${\cal N}=1^*$ geometry only a partial lift
exists \cite{pwn=1}. For these reasons  we will work mainly
with the Yang Mills$^*$ solution. Since the warp factor of the ${\cal N}=1^*$
theory behaves similarly to that of  Yang Mills$^*$ it is probable
that the two theories have qualitatively the same Wilson loop behaviour.

The 10d lift of YM$^*$ is
\begin{equation}
ds_{10}^2 = (\xi_+ \xi_-)^{1 \over 4} ds_{1,4}^2+
(\xi_+ \xi_-)^{-{3 \over 4}}  ds_{5}^2,
\end{equation}
where $ds_{1,4}^2$ is given in (\ref{metric4}) and
\begin{equation}
ds_5^2 =\xi_-  \cos^2 \alpha  ~ d\Omega_{+}^2
+ \xi_+ \sin^2 \alpha  ~ d\Omega_{-}^2
+\xi_+ \xi_- d\alpha^2.
\end{equation}
The functions $\xi_\pm$ are given by
\begin{equation}
\xi_\pm = c^2 \pm s^2 \cos 2\alpha,
\hspace{0.5cm} c= \cosh \lambda, \hspace{0.5cm}
s = \sinh \lambda.
\end{equation}

%The dilaton is given, in unitary gauge, by the functions
%\begin{equation}
%f = { 1 \over \xi^{1/2}} \sqrt{\cosh^2 \lambda + (\xi_+ \xi_-)^{1/2}
%\over 2}, \hspace{0.1cm}
%B = {  \sinh^2 \lambda \cos 2 \alpha \over  \cosh^2 \lambda
%+ (\xi_+ \xi_-)^{1/2}}
%\end{equation}
%In the more usual language the axion-dilaton field is given by
%\begin{equation}
%C + i e^{-\Phi} = i { ( 1 - B ) \over (1 + B)} = i \sqrt{\xi_- \over \xi_+}
%\end{equation}

The dilaton is given by
\begin{equation}
e^{-\Phi}  =  \sqrt{\xi_- \over \xi_+},
\end{equation}

\noindent The two-form potential is given by
\begin{equation}
A_{(2)}=i A_{+} \cos^3 \alpha \, \cos \theta_{+}
d \theta_{+} \wedge d \phi_{+}
-  A_{-}
\sin^3 \alpha \, \cos \theta_{-} d \theta_{-} \wedge d \phi_{-},
\label{bfield}
\end{equation}
with $A_{\pm}=\sinh 2\,\lambda / \xi_\pm$. Finally the four-form potential lifts to
\begin{equation}
F_{(4)}=  F +\star F, \hspace{0.5cm} F
= dx^{0}\wedge dx^{1}\wedge dx^{2}\wedge dx^{3}\wedge d\omega,
\end{equation}
where $\omega(r)=e^{4A(r)} A'(r)$.

We can calculate the Wilson loop behaviour by lying a string in an $x_{//}$
direction and letting it move in $r$ and $\alpha$. The action for
the string in Einstein frame is given by
\beq
S  = \frac{1}{2\pi \alpha'}\int d^2 \sigma e^{\Phi/2} \sqrt{{\rm det} G}.
\eeq
With this choice of orientation, the fundamental string does not
couple to the background B-field (\ref{bfield}) which has non-zero
components only on the internal manifold.
In the static gauge ($\sigma=x^1=x$, $\tau=x^0$) and using the
$z$ coordinates ($ds^2 = e^{2A(z)} (dx^2 + dz^2)$) we obtain
\beq
S =  \frac{T}{2\pi} \int d x  \; e^{2 A}
\sqrt{ \xi_+}  \sqrt{1 + \left({d z \over dx}\right)^2
+ e^{-2 A} \left({d \alpha \over dx}\right)^2}\label{ac}.
\eeq
where $T$ is the period resulting from the time integration.
The resulting equations of motion for $z$ and $\alpha$ are then
\beq
{d \over dx} \left[{e^{2 A}  \sqrt{ \xi_+} z^{'} \over
\sqrt{1 + z^{'2}
+ e^{-2 A} \alpha^{'2}}}\right] - {d \over d z} \left[ e^{2 A}
\sqrt{ \xi_+} \sqrt{1 + z^{'2}
+ e^{-2 A} \alpha^{'2}}\right]=0,
\eeq

\beq
{d \over dx} \left[{e^{2 A}  \sqrt{ \xi_+} \alpha^{'} \over
\sqrt{1 + z^{'2}
+ e^{-2 A} \alpha^{'2}}}\right] - {d \over d \alpha} \left[ e^{2 A}
\sqrt{ \xi_+} \sqrt{1 + z^{'2}
+ e^{-2 A} \alpha^{'2}}\right]=0.
\eeq

These can be solved numerically but there are an array of solutions.
The new feature relative to pure AdS is that the strings can have a
non-trivial $\alpha$ profile. The reason for this is that the fermion masses
we introduced break the SO(6) symmetry to $SO(3) \times SO(3)$ so there
are potentially different strings connecting particles in different subgroups
of the SO(6). To begin with lets concentrate on strings that are between two
identical particles and hence have no $\alpha$ variation.
There are such strings - in the $\alpha$ EoM the last potential term is
given by

\beq
 - {d \sqrt{ \xi_+} \over d \alpha} \left[ e^{2 A} \sqrt{1 + z^{'2}
+ e^{-2 A} \alpha^{'2}}\right] = - { \sinh^2 \lambda \sin 2 \alpha
\over \sqrt{\xi_+}}\left[ e^{2 A} \sqrt{1 + z^{'2} + e^{-2 A}
\alpha^{'2}}\right], \eeq so vanishes if $\alpha = n \pi/2$. The
kinetic term also vanishes if $\alpha =$constant so these are
solutions of the equations of motion.

As a simple case we study strings for which $\alpha = \pi/2$ which
implies $\xi_+=1$. The $z$ EoM then becomes

\beq {d \over dx} \left[{e^{2 A}   z^{'} \over \sqrt{1 +
z^{'2}}}\right] - {d (e^{2 A}) \over d z} \sqrt{1 + z^{'2}}=0.
\eeq

\noindent This is straightforward to solve numerically
in the background $A(z)$  appropriate to YM$^*$ (shown in Fig 2).

Consider initial conditions where we start the string at $z=-0.5$,
which is in the AdS like region. We then vary the derivative of
$z'$ and shoot off strings to more negative $z$ corresponding to
the interior of the deformed AdS space - we show numerical results
of this type in Fig 5. Initially as $z'$ increases the strings
penetrate the geometry  more and return to $z=-0.5$ at larger $x$
indicating that they describe a quark anti-quark pair that are
more widely separated. This is standard AdS behaviour. However
when the strings begin to enter the deformed space the behaviour
changes. At a critical value of $z'$ the string, although still
penetrating deeper into the space, returns to $z=-0.5$ at a {\it
shorter} quark separation. Thus there is a maximum quark
separation described by the Euler Lagrange equation solutions.

\begin{center}
\hskip-10pt{\lower15pt\hbox{ \epsfysize=2. truein
\epsfbox{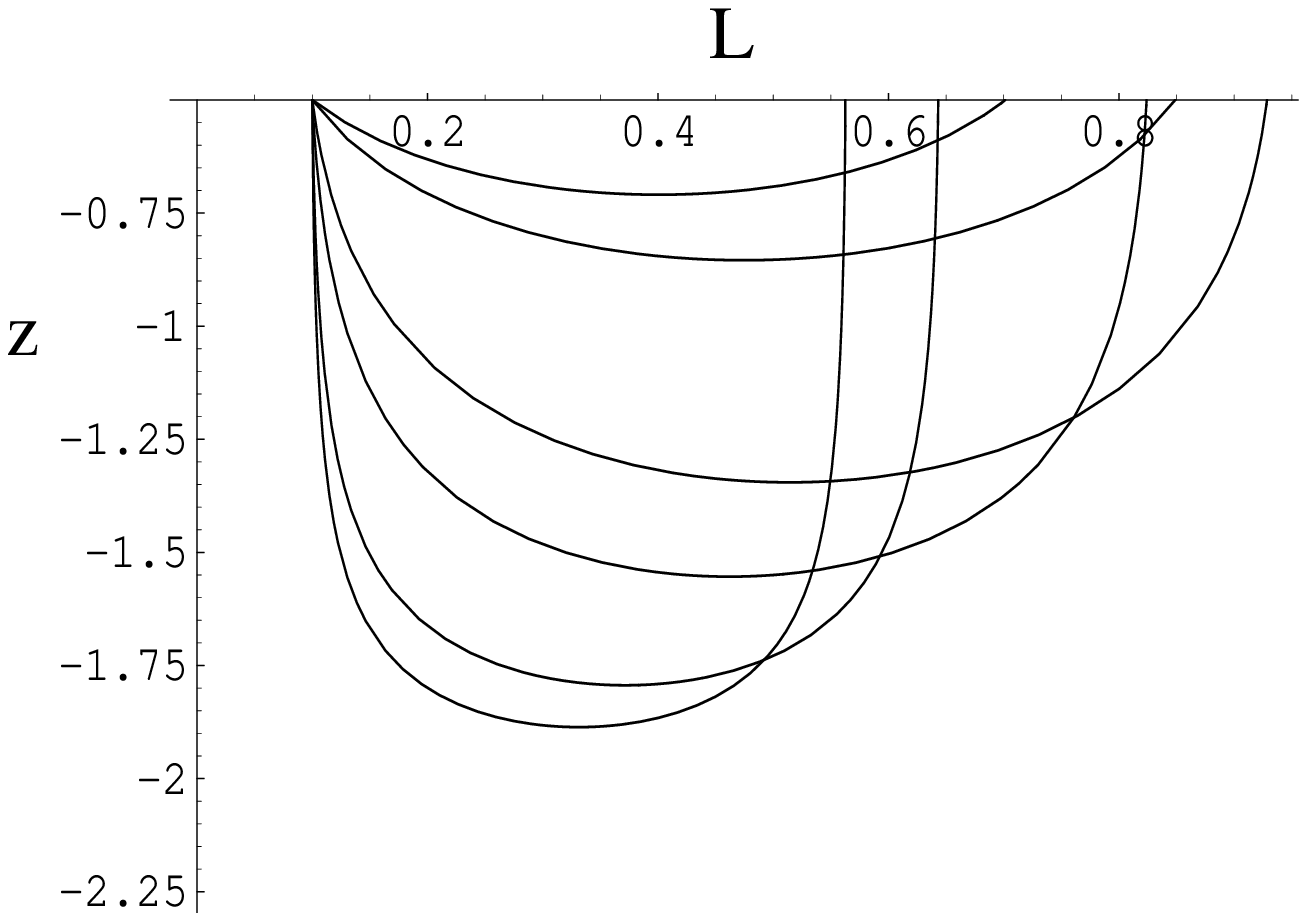}}}\medskip

Figure 5: Wilson loops in YM$^*$ showing how the depth the
probe string penetrates into the deformed space depends on the quark
separation (or equivalently the initial condition of $z'$).
\end{center}

To get a better handle on this behaviour we can use the $x$ independence
of the Lagrangian which implies the Hamiltonian is conserved
\beq
{\sqrt{\xi_+} e^{2A} \over \sqrt{1 + z^{'2}+\alpha^{'2} e^{-2 A}}}
= {\rm constant}.
\eeq
Thus, for $\alpha$ constant,
\beq
z' = \sqrt{{e^{4A} \xi_+  \over c^2} -1 }.
\eeq
We can determine the integration constant $c$ by evaluating the above
equation at the point of maximum depth reach by the string, $z_0$,
where $z'=0$ and which corresponds to $x=0$ (because of the
the symmetry of the problem)
%Positioning a string so that at $x=0$  $z=z_0$ and $z'=0$, we find:
\beq
c^2 = \xi_+(z_0) e^{4 A(z_0)}.
\eeq
The quark anti-quark separation is then given by
\beq
{L \over 2} = \int^{z_{max}}_{z_0} dz { 1 \over \sqrt{ {e^{4A}\xi_+  \over
c^2} - 1}}.
\eeq
The energy of the string is given by $S/T$
so we find
\beq
E = { 1 \over \pi c} \int^{z_{max}}_{z_0} dz { e^{4A}\xi_+  \over
\sqrt{ { e^{4A}\xi_+ \over c^2} - 1}}.
\eeq

These equations are again straightforward to solve numerically in
the YM$^*$ background. In Fig 6 we show plots of the quark
anti-quark separation vs $z_0$, the energy of the string vs $z_0$
and finally the energy of the string vs quark anti-quark
separation. We see again the maximum separation but now also that
the strings that penetrate into the interior of the deformed space
are higher action.

\begin{center}
\hskip-10pt{\lower15pt\hbox{ \epsfysize=1.5 truein
\epsfbox{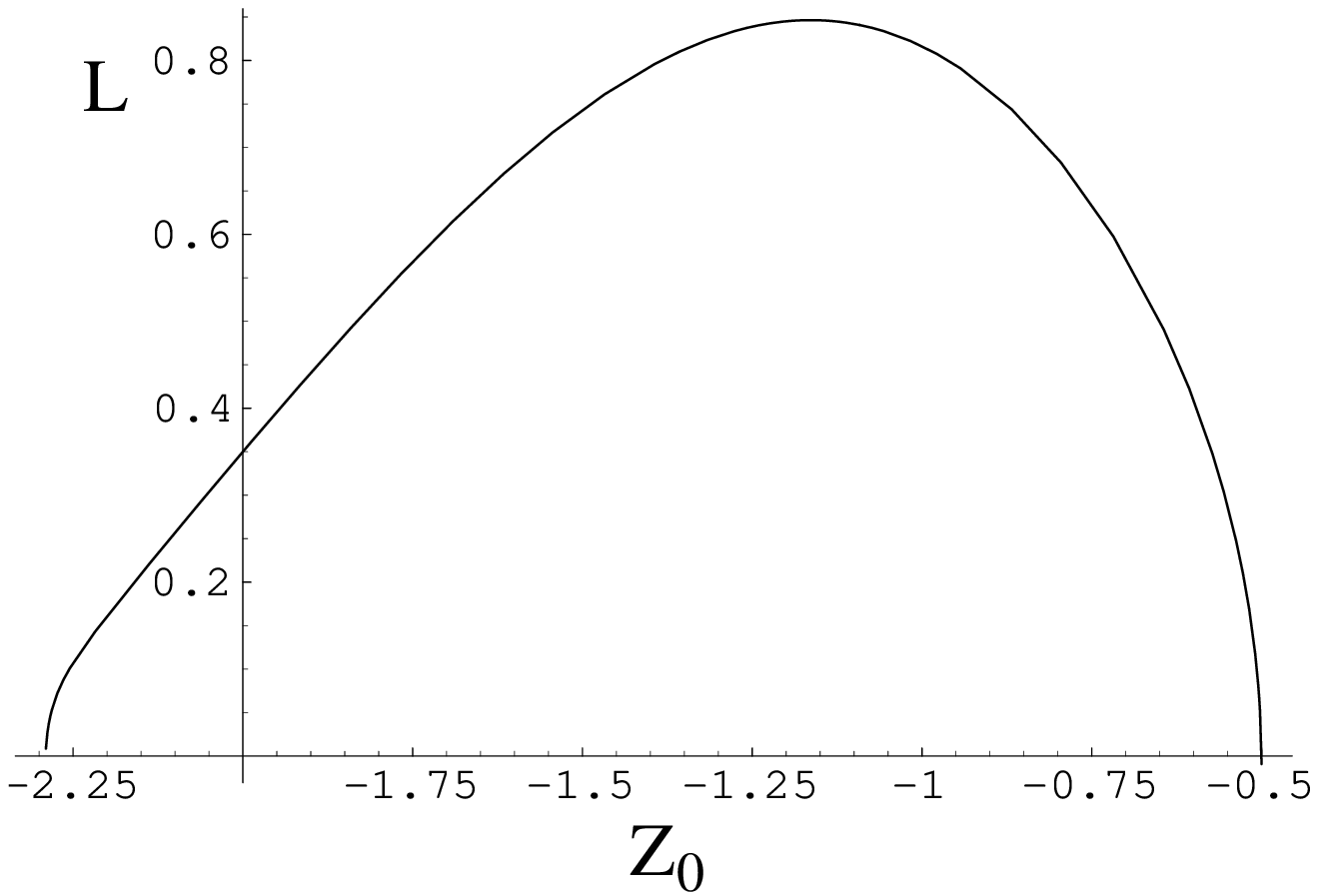}}} \hspace{0.25cm} \hskip-10pt{\lower15pt\hbox{
\epsfysize=1.5 truein \epsfbox{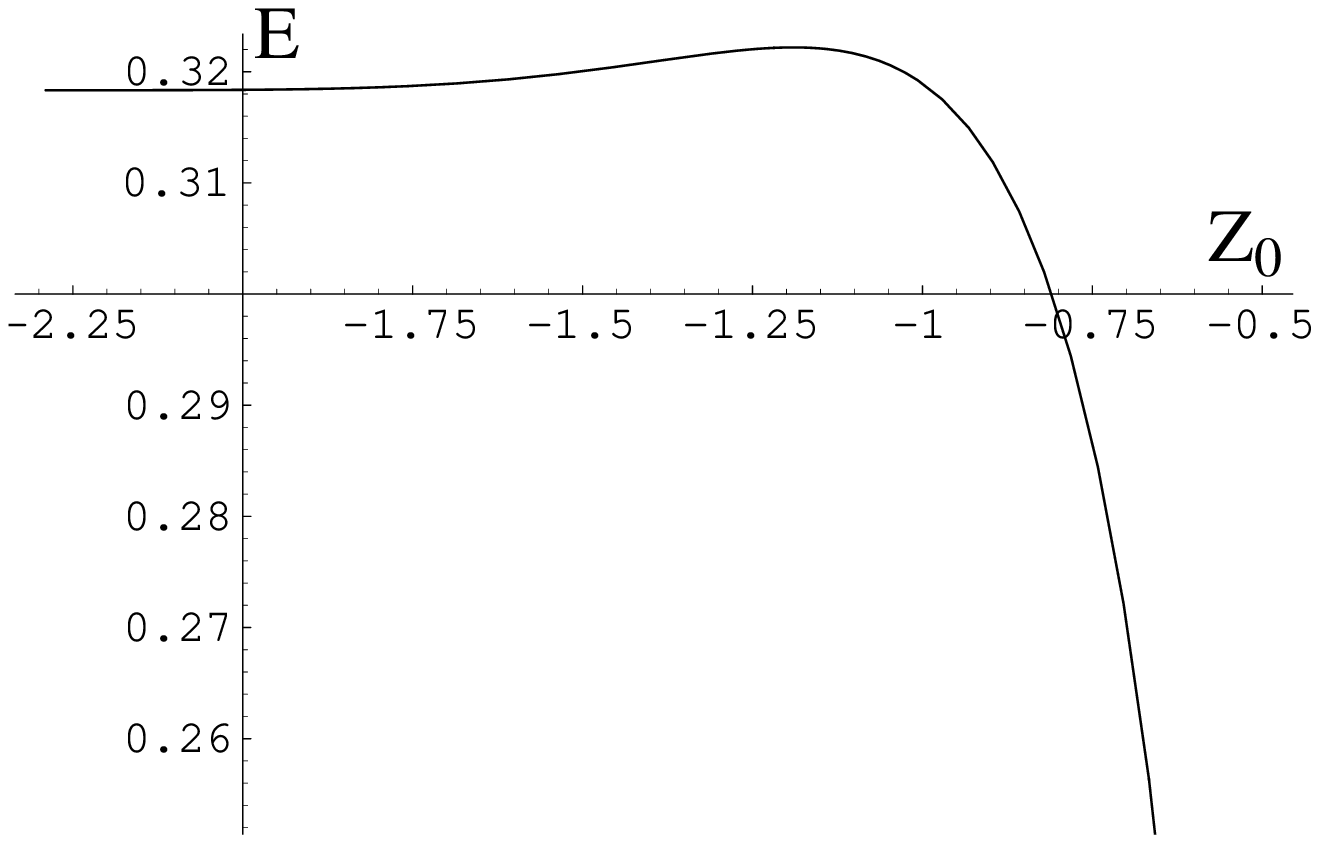}}} \hspace{0.25cm}
\hskip-10pt{\lower15pt\hbox{ \epsfysize=1.5 truein
\epsfbox{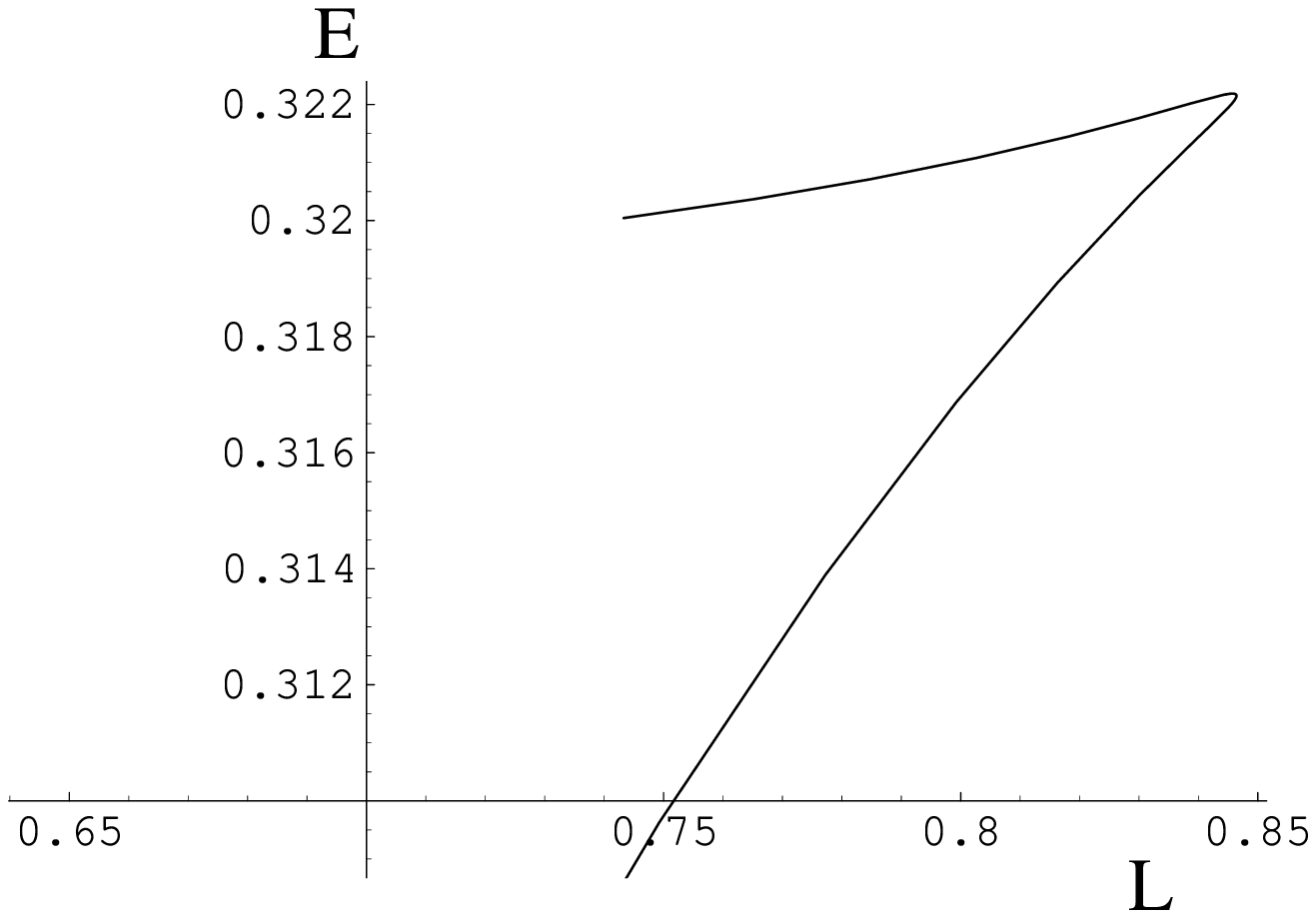}}}\medskip

Figure 6: YM$^*$ Wilson loop results at $\alpha=\pi/2$; we plot the
quark separation $L$ vs the maximum depth $z_0$ that the string
reaches; the energy of the string vs the maximum depth; and
finally the energy vs the quark separation.  \end{center}

This behaviour naively suggests string breaking by quark
anti-quark pair production (an interpretation we put forward in
the first version of this paper). We must be careful though since
the Euler Lagrange equations only return local turning points of
the solution (one of our solutions for fixed separation must
therefore be a maximum) and it is possible they miss some run-away
direction in the potential. To address this issue we should
consider a set of strings linking quarks at fixed separation but
which penetrate the space to different degrees. Ideally one should
include the true solution of the Euler Lagrange equations in this
set but we will learn all we need from a simpler set of
configurations. We take just a sinusoidal shape for the string

\beq z(x) =  -A \sin \pi x/L \eeq where we vary the amplitude $A$.

In Fig. 7 we plot the action of these configurations as a function
of $A$ for $L$ smaller than, larger than and close to the critical
quark separation found above.

\begin{center}
\hskip-10pt{\lower15pt\hbox{ \epsfysize=2.5 truein
\epsfbox{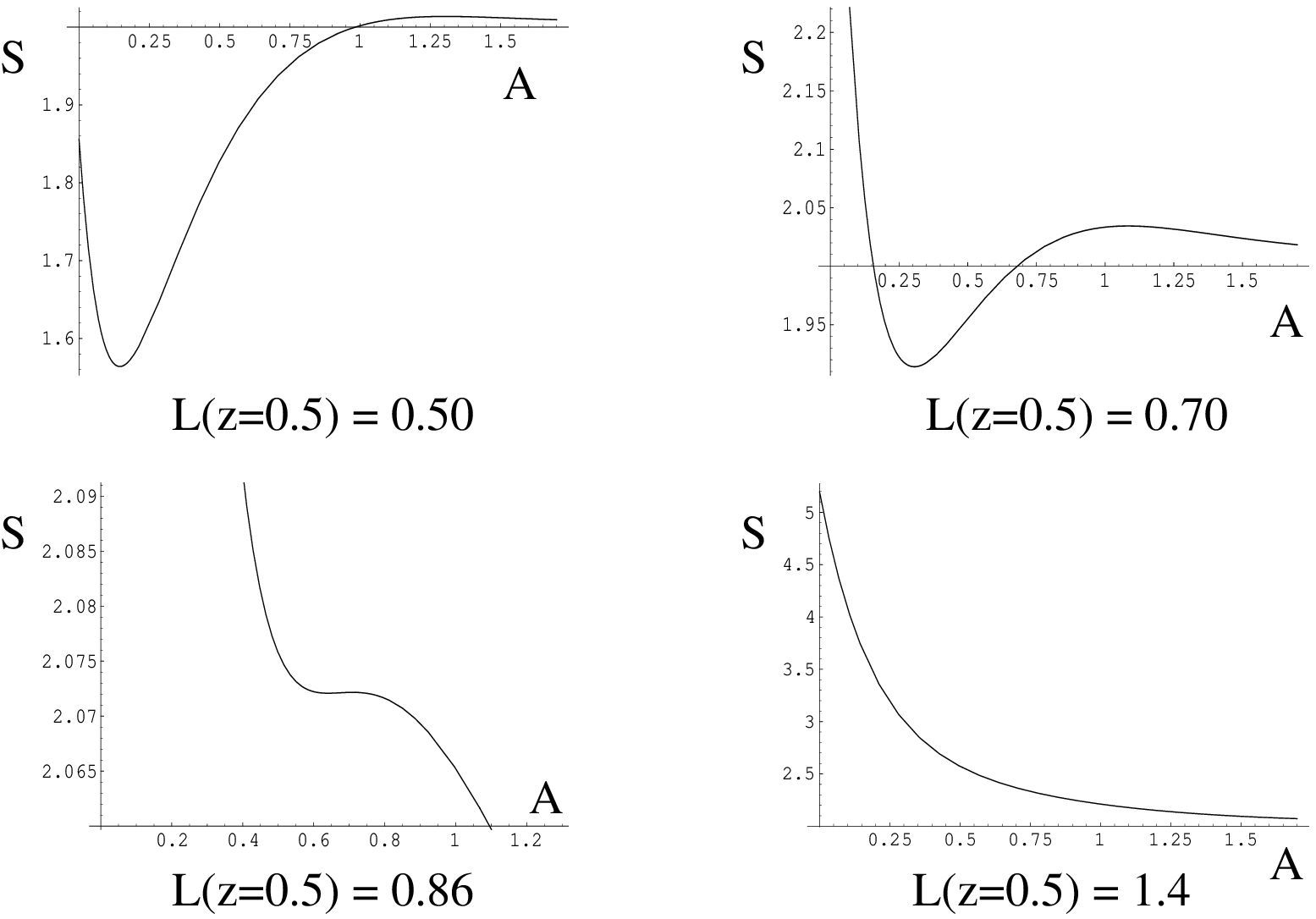}}} \medskip

Figure 7: The action of sinusoidal string configurations against
amplitude (penetration into the space) for four different values
of the quark separation $L$.
\end{center}

It is now clear what the Euler Lagrange equation solutions
correspond to. There is first a minimum of the action
corresponding to an AdS-like string, then a  maximum (the second
solution above) and finally the action falls as the string moves
into the singularity at $z=-2.3$. When the quarks are close the
global minimum is the AdS-like solution but at a critical
separation there is a  transition and the preferred configuration
is that which falls into the singularity. We have also looked at
solutions at fixed $\alpha=0$ and with varying $\alpha$. Although
there is a little more structure in the $\alpha$ variation the
same instability exists as that just described.

We conclude that we can not understand the behaviour of the Wilson
loops at the supergravity level for the IR. One could imagine that
with an appropriate stringy resolution of the singularity, perhaps
in terms of some fuzzy brane configuration, the situation might be
remedied. It could be that the strings prefer to form a bound
state with the interior branes and lie along the singularity (the
mechanism suggested by Polchinski and Strassler \cite{ps}) giving
rise to a confining potential but this must remain speculation in
these cases.

\section{Summary}

In this paper we have studied aspects of confinement in
two deformed AdS geometries, Yang Mills$^*$ and ${\cal
N}=1^*$.
We have shown that they describe discrete glueball spectra
indicating
that the geometries do confine. Study of bound states of
the massive fermions
in the geometries show that there is little
decoupling of these massive states.
Nevertheless the glueball spectra agree well with other
gravity duals with
the same infra-red degrees of freedom, even if the
ultra-violet of the theories
differ significantly.

We also studied the behaviour of Wilson loops in the 10d lift of
Yang Mills$^*$. We found though that when a quark pair is
separated beyond some critical separation the geometry prefers a
string configuration connecting the quarks that falls into the
singularity. It is therefore not possible to understand Wilson
loop behaviour without a stringy resolution of the interior which
currently is lacking.

\vspace{2cm}

\noindent {\bf Acknowledgements:} \\[2mm]
N.E. is grateful to PPARC for the support
of an Advanced Fellowship. D.C. is grateful to PPARC for the support
of studentships. M.P. is supported by the European Commission Marie
Curie Postdoctoral Fellowship under contract number
HPRN--CT--2001--01277 and partially by the EU contract
HPRN--CT--2000--00122. R.A. was supported by INFN and MURST and at a later stage by a PPARC postdoctoral fellowship.\\
R.A. would like to thank Prof. M. Bianchi and Prof. A. Zaffaroni for very useful discussions and
the physics department of Southampton University for kind hospitality.

\end{document}